\documentclass[aps,preprint]{revtex4}%
\usepackage{amsfonts}
\usepackage{amsmath}
\usepackage{amssymb}
\usepackage{graphicx}%
\begin{document}
\title{Teleportation with trapped ions in a magnetic field gradient }
\author{Z.J. Deng}
\email{dengzhijiao926@163.com}
\author{M. Feng}
\author{K.L. Gao}
\affiliation{State Key Laboratory of Magnetic Resonance and Atomic and Molecular Physics,
Wuhan Institute of Physics and Mathematics, Chinese Academy of Sciences, Wuhan
430071, China}
\affiliation{Center for Cold Atom Physics, Chinese Academy of Sciences, Wuhan 430071, China}
\affiliation{Graduate School of the Chinese Academy of Sciences, Beijing 100039, China}

\begin{abstract}
By means of the Ising terms generated by Coulomb interaction between ions and
the magnetic field gradient, we carry out teleportation with insurance with
trapped ions. We show the feasibility and the favorable feature of our scheme
by comparing with the recently achieved teleportation experiments with trapped ions.

\end{abstract}

\maketitle

\section{\textbf{INTRODUCTION}}

The trapped-ion system is one of the most promising candidates for quantum
information processing (QIP), which has achieved significant experimental
advances, for example, entanglement of few ions \cite{1}, realization of
simple quantum gates or quantum algorithms \cite{2,3,4} and so on.
Particularly, the recently achieved teleportation \cite{5,6} in ion traps
paves the way towards scalable QIP with trapped ions.

\ Much effort has been paying on overcoming barriers of scalability and
fidelity, such as decoherence due to heating, imprecise laser focusing and
instability of operation at optical frequencies \cite{7}. We have noticed some
technical improvement in experiments as well as some potentially practical
proposals. In this Letter, we focus on a long wave-length scheme, which
corresponds to a modified ion trap with a magnetic field gradient applied to
the confined ions \cite{8,9,10}. In this scheme, the qubits are encoded in the
atomic hyperfine levels, and due to the magnetic field gradient the ions can
be distinguished in frequency space and individually addressed by microwaves.
This new scheme is advantageous over those in original ion traps in following
points: (1) Raman-type radiation is not necessary. So the qubit rotation
carried out by microwaves is straightforward. (2) Ising terms due to the
magnetic field gradient give the possibility of quantum gating by the well
developed NMR-type implementation. (3) Quantum computing can be made with an
array of ions confined in either a linear trap or some individual micro-traps,
where the spin-spin coupling strength between different ions is adjustable in
the latter case. (4) The cheap source of microwave makes experiments performed
more easily than lasers.

\ In this Letter, we study how to carry out teleportation based on the models
in Refs. \cite{8,9,10}. Our calculation starts from the multi-trap model in
\cite{10}. Since three ions confined in a linear trap can be considered
mathematically as a special situation of the multi-trap model, i.e., the same
trap frequency for each of the ions, our discussion will cover the three-ion
cases in either a single trap or three traps respectively. As shown in Fig.1,
we first consider three $^{171}$Yb$^{+}$ ions, each confined\ in an individual
micro-trap. The hyperfine levels ($|1\rangle=$%
$\vert$%
F=1, m$_{F}$=+1%
$>$
and $|0\rangle=$%
$\vert$%
F=0%
$>$%
) with qubits encoded are split differently for different ions. To avoid
heating, we will carry out teleportation only by carrier transitions by using
the method in NMR quantum computation. Comparing with the achieved
teleportation experiments, we show the merits of our scheme and discuss
whether it could be achieved by current techniques.

\bigskip

\section{\textbf{THE CONFIGURATION OF THREE IONS}}

For a system with three ions in Fig. 1, the coupling strength between
different ions can be adjusted by choosing the position and the trap frequency
of each trap appropriately. According to Ref. \cite{10},\textbf{\ }we have in
units of $\hbar=1$,

\
\begin{equation}
H_{0}=\overset{3}{\underset{i=1}{\sum}}\frac{1}{2}w_{i}(z_{0,i})\sigma
_{z,i}-\frac{1}{2}J\sigma_{z,1}\sigma_{z,2}-\frac{1}{2}J\sigma_{z,2}%
\sigma_{z,3}-\frac{1}{2}J_{13}\sigma_{z,1}\sigma_{z,3}+\underset{\ell
=1}{\overset{3}{\sum}}\nu_{\ell}a_{\ell}^{+}a_{\ell},\ \label{1}%
\end{equation}

where\ \ $J_{ij}=\overset{3}{\underset{\ell=1}{\sum}}\frac{\hbar}{2m\nu_{\ell
}^{2}}D_{i,\ell}D_{j,\ell}\frac{\partial w_{i}}{\partial z}|_{z_{0,i}}%
\frac{\partial w_{j}}{\partial z}|_{z_{0,j}}$ with $D$ a unitary matrix
related to the Hessian of the potential, $w_{i}(z_{0,i})$ is the
position-dependent transition frequency for ion i in the magnetic field
gradient with $z_{0,i}$ the equilibrium position, $\nu_{\ell}$ is the $\ell$th
vibrational frequency, and $\sigma_{zi}$ is the usual Pauli operator for ion
i. We have adjusted $J_{12}=J_{23}=J$\ by simply setting the trap frequencies
of micro-traps for ions 1 and 3 to be equal. To avoid exciting the vibrational
mode, we will only consider the carrier transition. So we neglect the last
term in Eq. (1). Direct algebra shows that in the space spanned by
$|0\rangle_{1}|0\rangle_{2}|0\rangle_{3},$ $|1\rangle_{1}|0\rangle
_{2}|0\rangle_{3},$ $|0\rangle_{1}|1\rangle_{2}|0\rangle_{3},$ $|0\rangle
_{1}|0\rangle_{2}|1\rangle_{3},$ $|1\rangle_{1}|1\rangle_{2}|0\rangle_{3},$
$|1\rangle_{1}|0\rangle_{2}|1\rangle_{3},$ $|0\rangle_{1}|1\rangle
_{2}|1\rangle_{3},$ $|1\rangle_{1}|1\rangle_{2}|1\rangle_{3},$ the
eigenenergies are $-\frac{w_{1}}{2}-\frac{w_{2}}{2}-\frac{w_{3}}{2}%
-J-\frac{J_{13}}{2},$ $\ \frac{w_{1}}{2}-\frac{w_{2}}{2}-\frac{w_{3}}{2}%
+\frac{J_{13}}{2},$ $\ -\frac{w_{1}}{2}+\frac{w_{2}}{2}-\frac{w_{3}}%
{2}+J-\frac{J_{13}}{2},$ $\ -\frac{w_{1}}{2}-\frac{w_{2}}{2}+\frac{w_{3}}%
{2}+\frac{J_{13}}{2},$ $\ \frac{w_{1}}{2}+\frac{w_{2}}{2}-\frac{w_{3}}%
{2}+\frac{J_{13}}{2},$ $\ \frac{w_{1}}{2}-\frac{w_{2}}{2}+\frac{w_{3}}%
{2}+J-\frac{J_{13}}{2},$ $\ -\frac{w_{1}}{2}+\frac{w_{2}}{2}+\frac{w_{3}}%
{2}+\frac{J_{13}}{2},$\ $\ \frac{w_{1}}{2}+\frac{w_{2}}{2}+\frac{w_{3}}{2}-J-$
$\frac{J_{13}}{2}$ ,\ respectively. As shown in Fig. 2, the carrier transition
frequencies for each ions are dependent on the states of the other two ions
\cite{11}.

In Table I, under the restriction $\varepsilon_{\max}<0.05$($\varepsilon
_{\max}$ is the maximum of all $\varepsilon_{i,\ell}$ defined in the next
section), we list different cases in our consideration, in which the trap
frequencies and magentic field gradient are chosen for getting the biggest $J$
with respect to a certain neighboring trap distance $d$. It's obvious that the
smaller the neighboring trap distance $d,$ the bigger the $J$ obtainable.
Moreover, to resonantly excite a certain ion irrespective of the states of the
other two ions, we should have a microwave with bandwidth larger than the
maximum difference between the carrier transition frequencies regarding this
ion. In what follows, we will use the numbers in Table I regarding $d$=4$\mu
m$ to demonstrate our scheme. If the microwave bandwidth is $63.7\times2\pi
kHz,$ as we can see in Fig.2, we cannot distinguish different carrier
transitions with this microwave for individual ions. Straightforward
calculation also shows that the three vibrational frequencies are
approximately $1.32\times2\pi MHz,1.54\times2\pi MHz,1.70\times2\pi MHz,$ and
the resonance frequency shift between neighboring qubits is $g\mu_{B}%
\frac{\partial B}{\hbar\partial z}(d+\Delta)\approx64.8\times2\pi MHz$
($\Delta$ represents the distance between equilibrium position of ion 1 or ion
3 and their own trap centers.)

\section{\textbf{\ CONTROLLED-NOT GATE}}

The controlled-NOT (CNOT) gate is the most commonly used two-qubit quantum
gate. When a microwave is added, the interaction Hamiltonian in our system
differs from that in original ion trap only by replacing the Lamb-Dicke
parameter $\eta_{i,\ell}$ with $\eta_{i,\ell}^{^{\prime}}=$ $\sqrt
{\eta_{i,\ell}^{2}+\varepsilon_{i,\ell}^{2}}$, where the subscript i denotes
ion i, $\ell$ is for the $\ell$th collective motional mode, $\eta_{i,\ell}$ is
about 10$^{-6}$, and $\varepsilon_{i,\ell}=D_{i,\ell}(\sqrt{\hbar/2m\nu_{\ell
}}\frac{\partial w_{i}}{\partial z}|_{z_{0,i}})/\nu_{\ell}$ is an additional
Lamb-Dicke parameter due to the magnetic field gradient \cite{9}. In the
Paschen-Bach limit ($B_{0}\sim1$), as the frequency gradients are independent
of z, $\frac{\partial w_{i}}{\partial z}|_{z_{0,i}}=\frac{2}{\hbar}\mu
_{B}\frac{\partial B}{\partial z}$\cite{10}$.$Given the numbers in Table I for
d=4$\mu$m, the maximum $\varepsilon_{i,\ell}$ is about $0.0340$. So all the
$\eta_{i,\ell}^{^{\prime}}$ are much smaller than 1, and we can get to a
reduced unitary evolution operator for addressing ion i by a microwave in the
interation picture,

$\ \ \ \ \ \ \ \ \ \ \ \ \ \ $%
\begin{equation}
U_{I}^{i}(\theta,\phi)=\exp[i\frac{\theta}{2}(e^{-i\phi}\sigma_{+}+e^{i\phi
}\sigma_{-})],\ \ \label{2}%
\end{equation}

\ where $\theta=\Omega t$ with $\Omega$ the Rabi frequency, and $\phi$\ is
related to the position of the ion i in the microwave. We suppose $\Omega$ to
be of the order of MHz. If $\phi=0$, then $U_{I}=\exp(i\frac{\theta}{2}%
\sigma_{x})$, and if $\phi=\frac{\pi}{2},$ then we have $U_{I}=\exp
(i\frac{\theta}{2}\sigma_{y}).$

In NMR quantum computation, CNOT gate is realized by a sequence of pulses as
below \cite{12},

$\ $%
\begin{equation}
U_{cnot}(i,j)=e^{-i\pi/4}e^{-i\pi/4\sigma_{y,j}}e^{+i\pi/4\sigma_{z,i}%
}e^{+i\pi/4\sigma_{z,j}}e^{-i\pi/4\sigma_{z,i}\sigma_{z,j}}e^{+i\pi
/4\sigma_{y,j}}, \label{3}%
\end{equation}

where ions i and j act as control and target qubits, respectivley.
$e^{-i\pi/4\sigma_{z,i}\sigma_{z,j}}$ can be realized by means of the coupling
term $-\frac{1}{2}J\sigma_{z,i}\sigma_{z,j}$ in $H_{0}$ and refocusing
techniques must be used to eliminate undesired evolution brought by the other
terms in $H_{0}$ \cite{13}. All the other terms in Eq. (3)\ can be implemented
by microwave pulses. But we note that, in the context of NMR, Rabi
frequency\ is much larger than other characteristic frequencies. In our case,
however, $w_{i}(z_{0,i})$\ $\sim13\times2\pi$GHz is the biggest frequency,
much bigger than Rabi frequency. In order to avoid undesired evolution due to
$w_{i}(z_{0,i}),$ we need carefully control the pulse length $T$\ to satisfy
\begin{equation}
w_{i}(z_{0,i})T=2\pi N(i=1,2,3;N=1,2,3....),\text{ \ \ \ } \label{4}%
\end{equation}
\ 

which can be done by properly adjusting $B_{0}$, $\partial B/\partial z$\ and
$\Omega$. In Table II, we give an example of $U_{cnot}(2,3)$. The total time
for completing a CNOT gate is about $3.84ms.$ If we decrease the neighboring
trap distance $d,$ the CNOT gating time can be shortened due to the
enlarged$\ J$.

\section{\ \textbf{TELEPORTATION}}

\ Teleportation \cite{14} is a disembodied transport of unknown quantum states
by combining quantum and classical channels. The two separated partners Alice
and Bob share two entangled particles 2 and 3 initially. After Alice inputs
her information by making joint measurement on particles 1 and 2, Bob is able
to recover the state of particle 1 in particle 3, assistent with some unitary
transformations conditional on the classical communication with Alice. Since
its proposal, teleportation has been experimentally realized in both
macroscopic \cite{15} and microscopic distances \cite{5,6,16}. In what
follows, we show how to carry out a teleportation in our system.

Considering ion 1 initially in an arbitrary state $\alpha|0\rangle_{1}%
+\beta|1\rangle_{1},$ ions 2 and 3 in states $\frac{1}{\sqrt{2}}(|0\rangle
_{2}+|1\rangle_{2})$ and $|1\rangle_{3}$\ respectively, after the CNOT gate
$U_{cnot}(2,3),$ we get the entangled state $\frac{1}{\sqrt{2}}(|0\rangle
_{2}|1\rangle_{3}+|1\rangle_{2}|0\rangle_{3})$ whose lifetime can be longer
than 100ms \cite{5}$.$ So at this stage the three ions are in a state

$\ \ \ \ \ \ \ \ \ \ \ $%
\begin{equation}
\text{ }|\Psi_{1}\rangle=\frac{1}{\sqrt{2}}(\alpha|0\rangle_{1}+\beta
|1\rangle_{1})(|0\rangle_{2}|1\rangle_{3}+|1\rangle_{2}|0\rangle_{3}).\text{ }
\label{5}%
\end{equation}

With another c-not gate $U_{cnot}(1,2),$ we have

\ \ \ \ \ \ \ \
\begin{equation}
\ |\Psi_{2}\rangle=\frac{1}{\sqrt{2}}[\alpha|0\rangle_{1}(|0\rangle
_{2}|1\rangle_{3}+|1\rangle_{2}|0\rangle_{3})+\beta|1\rangle_{1}(|1\rangle
_{2}|1\rangle_{3}+|0\rangle_{2}|0\rangle_{3})]. \label{6}%
\end{equation}

Then a Hadamard transformation $|0\rangle\longrightarrow\frac{1}{\sqrt{2}%
}(|0\rangle+|1\rangle)$ and $|1\rangle\longrightarrow\frac{1}{\sqrt{2}%
}(|0\rangle-|1\rangle)$ on ion 1 would yield

\ \ \ \ \ \ \ \
\begin{align}
\ |\Psi_{3}\rangle &  =\frac{1}{2}[|0\rangle_{1}|0\rangle_{2}(\alpha
|1\rangle_{3}+\beta|0\rangle_{3})+|0\rangle_{1}|1\rangle_{2}(\alpha
|0\rangle_{3}+\beta|1\rangle_{3})\label{7}\\
&  +|1\rangle_{1}|0\rangle_{2}(\alpha|1\rangle_{3}-\beta|0\rangle
_{3})+|1\rangle_{1}|1\rangle_{2}(\alpha|0\rangle_{3}-\beta|1\rangle
_{3})].\text{ \ \ \ }\nonumber
\end{align}

$\ \ \ \ \ \ \ \ \ \ \ \ \ \ \ \ $

Therefore, similar to Refs. \cite{5,6}, by making measurement on ions 1 and 2
in the bases $\{|0\rangle_{1}|0\rangle_{2},|0\rangle_{1}|1\rangle
_{2},|1\rangle_{1}|0\rangle_{2},|1\rangle_{1}|1\rangle_{2}\},$ we can obtain
$\alpha|0\rangle_{3}+\beta|1\rangle_{3}$\ assistent with appropriate unitary
qubit rotation $\sigma_{x},$ $I,$ $i\sigma_{y,}\ \sigma_{z}$. All these
operations can be achieved by microwave pulses by using Eq. (2). Because all
the four bases can be distinguished definitely, our teleportation is done with
insurance as in Refs. \cite{5,6}.

\bigskip

\section{\textbf{DISCUSSION AND CONCLUSION}}

Since the three ions under our consideration are confined in separate
micro-traps, our calculation above is actually for a multi-trap system.
However, different from the multiplexed traps in Ref. \cite{6}, the magnetic
field gradient in our case provides spin-spin couplings between the ions. So
we can entangle the ions without using vibrational mode as bus qubit or
detecting the leaking photons \cite{17}.

As the ions are distinguished in frequency space under the magnetic field
gradient, we can carry out teleportation without moving or hiding any ions,
which is much simpler than in Refs. \cite{5,6}. Due to the spin-spin coupling
in our system, we accomplish quantum gates by the way similar to NMR quantum
computing. This is different from the Cirac-Zoller CNOT gate in Ref. \cite{5},
which has to excite the vibrational mode, and is also different from the
geometric phase gating in Ref. \cite{6}. Although in the latter work the
geometric phase gate was made by only virtually exciting the vibrational mode,
like we are doing here, the movement of the ions between different traps,
which inevitably yields heating, reduces the fidelity of the teleported state.
Besides, the required bichromatic radiation in Ref. \cite{6}, based on the
Raman process, makes the experiment very challenging. In contrast, our method,
without ions' moving and by using the sophisticated microwave pulse sequenses,
is more straightforward.

However, the teleportation time in our case is longer than those in Refs.
\cite{5,6} by approximately two times, which is not good in view of
decoherence. To reduce the implementation time, we have to increase the Ising
interacion J, which can be done by reducing the inter-ion spacing d or
enlarging the magnetic field gradient. However, the smaller the inter-ion
distance, the more challenging the experimental implementation, i.e., more
focused laser beams or higher quality micro-traps are needed. Besides, the
bigger magnetic field gradient would increase the effective Lamb-Dicke
parameter. Since the Zeeman splitting of the ions are always changing due to
the vibration of the ions in the magnetic field gradient, to reduce
infidelity, we require the ions keeping strictly within the Lamb-Dicke limit
throughout the our scheme.

Although we neglect the vibrational modes, since the Ising term is from the
virtually coupling to the vibrational modes, we still need to consider heating
of the vibrational mode in our two-qubit gating. We know the heating time of
the ground vibrational mode to be of the order of 4 ms for a trap of the size
of \textbf{100}$\mu$m \cite{6}. Since the heating rate scales with the trap
size R as R$^{-4}$ \cite{18}, the micro-trap in our considerartion, which is
of the size of 4 $\mu$m, would own a heating time of the order of nanosec.
Therefore our multi-trap scheme is not achievable with current technique.
Nevertheless, if we turn to a linear trap with three ions confined, which
corresponds to the treatment in Refs. \cite{8,9} and also to the situation in
Ref. \cite{5} but with a magnetic field gradient applied, we find the
feasibility of our scheme. Based on Table III, we obtain the CNOT gating time
is about 4.94 ms for the three ions with the neighboring spacing of 4 $\mu$m.
Suppose in our case the heating time to be 1 ms, and only 10\% vibrational
mode being actually excited in our two-qubit gate, the actual heating time
would be 10 ms, long enough for our scheme.

Experimentally, three ions in a linear trap with the spacing of 4-5 $\mu$m has
been achieved \cite{5}, and the microwave with a certain bandwidth and a rapid
change of phase $\phi$\ is a sophisticated technique. Moreover, magnetic field
gradient up to $8000$ T/m is experimentally available in the near future
\cite{8}. We should mention here that in this long wave-length model lasers
are still necessary to provide the initially cooling of ions, initial state
preparation and final read-out, as mentioned in \cite{8,9,10}. Therefore, the
ions are required to be distingushed in the optical frequency domain, which
can be achieved by beam-focusing technique \cite{3,5}.

Compared with the teleportation by photons \cite{19}, our implementation is
deterministic, and the trapped ions are better candidates for storing
information than the flying photons. Moreover, although we make use of the
NMR-type inplementation, our gating is really performed on individual quantum
states, instead of spin ensemble in NMR system \cite{16}. So what we show here
is really a teleportation of a quantum state. Furthermore, we noticed the
Ising coupling widely used in solid state QIP, such as in semiconductor
quantum dots \cite{20}, in doped fullerenes \cite{11,21}, and so on. Quantum
gating in these solid-state systems is still very challenging experimentally
\cite{22}. Therefore, the relatively easier achievement of the CNOT and
teleportation in our trapped ions system would provide a possibility to test
different QIP proposals for above solid-state systems.

In summary, we have specifically investigated a teleportation scheme with
three trapped ions in a magnetic field gradient. We show the possibility to
realize teleportation without moving any ions and without exciting vibrational
modes. By considering currently available ion traps, we argue that our scheme
is achievable in a linear trap and will also be feasible in multi-trap devices
in the future if the heating problem can be overcame. Although the different
energy splittings of the ions in our system due to magnetic field gradient
require more attention to the refocusing, we don't think that those operations
would be more technically difficult than in original ion traps, as long as we
know exactly the energy splittings of each qubits and the ions are strictly
restricted within the Lamb-Dicke limit.

\bigskip

{\Large ACKNOWLEDGMENTS\ }

The authors acknowledge thankfully the helpful discussion with Derek Mc Hugh,
Jun Luo and Christof Wunderlich. MF is also grateful to Jason Twamley for his
hospitality and encouragement when he stayed in NUIM. Z.J. D is thankful for
warmhearted help from Guilong Huang, Baolong Lu, Daxiu Wei and Xiwen Zhu. This
work is supported by National Natural Science Foundation of China under
contract numbers 10474118 and 10274093.

\bigskip

\textbf{Note added \ \ }After finishing this Letter, we\textbf{ }become aware
of a work \cite{23}, in which the Ising coupling between ions is obtained by
changing the laser intensities and the polarizations. As it is mathematically
identical to the models in \cite{8,9,10}, our scheme can be in principle
applied to it.

\bigskip\newpage

\bigskip\ {\LARGE Figure Captions}

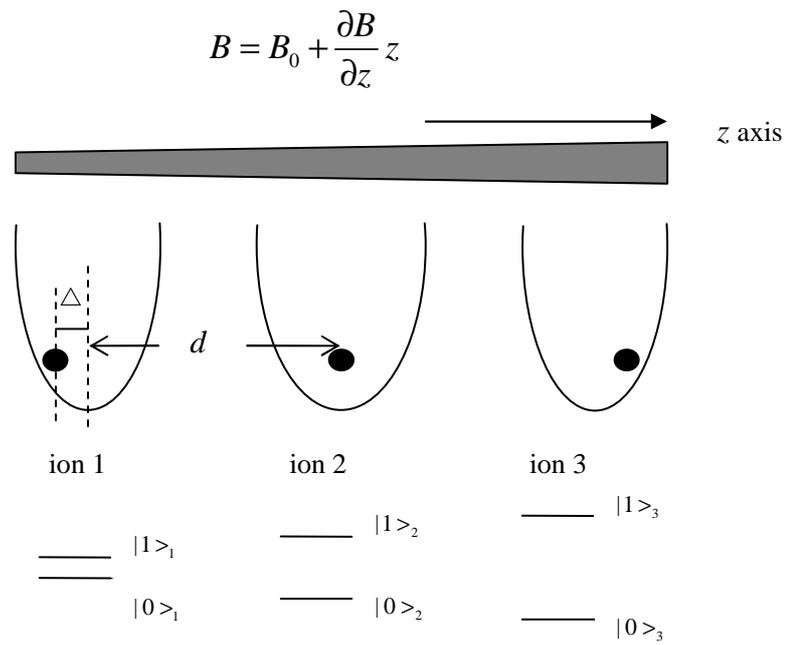
Fig. 1. Schematic diagram for three $^{171}$Yb$^{+}$ ions, each confined in an
individual micro-trap. The magnetic field gradient is applied along the $z$
axis (i.e. the ion array). So the energy differences of the ions are
position-dependent. $d$ is the separation between neighboring traps and
$\triangle$ denotes the deviation of the equilibrium position for ion 1 or ion
3 from their respective trap center.

\bigskip

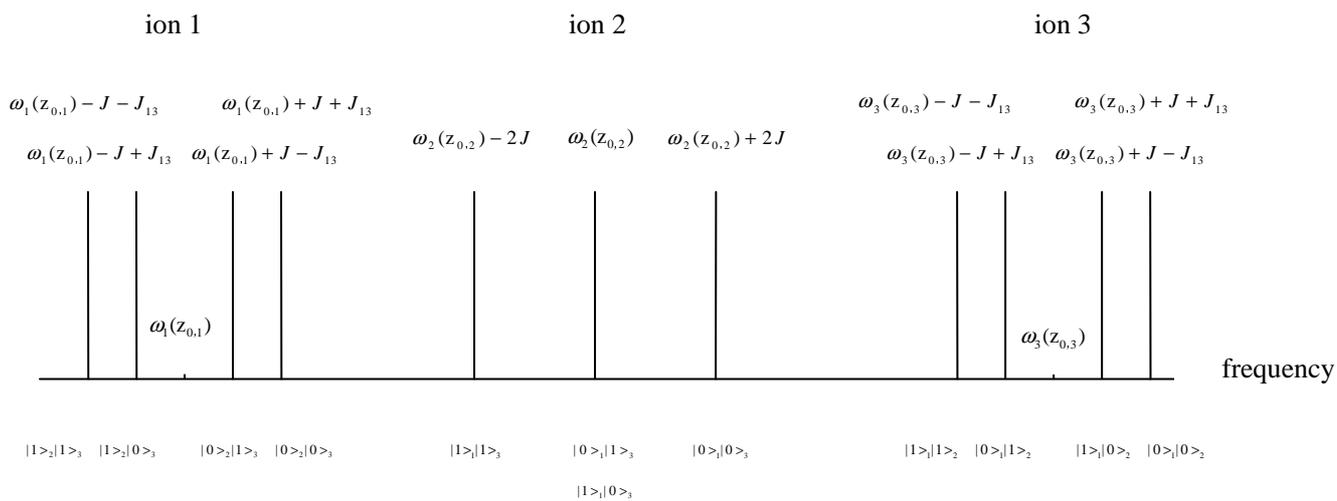
Fig. 2. Spectrum of the carrier transition frequencies, where each vertical
line represents a carrier transition frequency of a certain ion with
corresponding states of the other two ions labeled below.

\bigskip

\newpage\ 

\textbf{Table I.} Cases of three $^{171}$Yb$^{+}$ having the largest $J$ with
respect to a certain neighboring trap distance $d$, where $W_{1}$ is the trap
frequency for ion 1 and ion 3, $W_{2}$ is the trap frequency for ion 2,
$\partial B/\partial z$ denotes magnetic field gradient and $\Delta$
represents the distance between equilibrium position of ion 1 or ion 3 and
their own trap center and $h$ is the inter-ion distance.%

\begin{tabular}
[c]{|c|c|c|c|c|c|c|c|c|}\hline
$d(\mu m)$ & $\ W_{1}(MHz)$ & $\ \ W_{2}(MHz)$ & $\partial B/\partial z(T/m)$
& $\Delta(\mu m)$ & $\ \varepsilon_{\max}$ & $h(\mu m)$ & $\ J(kHz)$ &
$\ J_{13}(kHz)$\\\hline
$1$ & $3.20\times2\pi$ & $\ 0.097\times2\pi$ & $\ 1200$ & $0.779$ & $0.0376$ &
$1.779$ & $1.60\times2\pi$ & $\ 0.746\times2\pi$\\\hline
$2$ & $2.72\times2\pi$ & $\ 1.27\times2\pi$ & $\ 1000$ & $0.531$ & $0.0422$ &
$2.531$ & $1.07\times2\pi$ & $\ 0.337\times2\pi$\\\hline
$3$ & $1.85\times2\pi$ & $\ 1.10\times2\pi$ & $\ \ 600$ & $0.578$ & $0.0427$ &
$3.578$ & $0.645\times2\pi$ & $\ 0.179\times2\pi$\\\hline
$4$ & $1.37\times2\pi$ & $\ 1.24\times2\pi$ & $\ \ 500$ & $0.628$ & $0.0340$ &
$4.628$ & $0.459\times2\pi$ & $0.135\times2\pi$\\\hline
$5$ & $1.21\times2\pi$ & $\ 1.05\times2\pi$ & $\ \ 400$ & $0.558$ & $0.0382$ &
$5.558$ & $0.334\times2\pi$ & $0.0820\times2\pi$\\\hline
$6$ & $0.971\times2\pi$ & $\ 0.891\times2\pi$ & $\ \ 300$ & $0.612$ & $0.0356$
& $6.612$ & $0.254\times2\pi$ & $0.0623\times2\pi$\\\hline
$7$ & $0.732\times2\pi$ & $\ 0.700\times2\pi$ & $\ \ 200$ & $0.777$ & $0.0319$
& $7.777$ & $0.197\times2\pi$ & $0.0515\times2\pi$\\\hline
\end{tabular}

\bigskip

\bigskip

\textbf{Table II.} The implementation pulses and time for each term of
$U_{cnot}(2,3)$, where $U_{0}$ denotes the corresponding unitary evolution
operator of $H_{0}$ (without including the last term in Eq. (1)), $X_{i}^{2}$
means $U_{I}^{i}(\pi,0)$, and we adopt the numbers in Table I for d=4$\mu$m.
For simplicity, we assume the implementation time for any single qubit
rotation $U_{I}^{i}(\theta,\phi)$ to be $t_{m}=2.5\mu s.$

$\ $%
\begin{tabular}
[c]{|c|c|c|c|c|}\hline
\ term & implementation \ \ \ \ pulses & \ time & \ estimated time & $%
\begin{array}
[c]{c}%
\text{total estemated}\\
\text{time}%
\end{array}
$\\\hline
$e^{+i\pi/4\sigma_{y,3}}$ & $U_{I}^{3}(\frac{\pi}{2},\frac{\pi}{2}%
)$\textbf{\ } & $t_{m}$ & $\ \ 2.5\mu s$ & \\\cline{1-4}%
$e^{-i\pi/4\sigma_{z,2}\sigma_{z,3}}$ & $%
\begin{array}
[c]{c}%
X_{3}^{2}X_{2}^{2}U_{0}(\frac{t}{4})X_{1}^{2}U_{0}(\frac{t}{4})X_{3}^{2}%
X_{2}^{2}\\
U_{0}(\frac{t}{4})X_{1}^{2}U_{0}(\frac{t}{4})\mathbf{(t=\frac{7\pi}{2J})}%
\end{array}
$ & $t+4t_{m}$ & $\ \ 3.82ms$ & $3.84ms$\\\cline{1-4}%
$e^{+i\pi/4\sigma_{z,3}}$ & $\ \ \ U_{I}^{3}(\frac{\pi}{2},\frac{\pi}{2}%
)U_{I}^{3}(\frac{\pi}{2},0)U_{I}^{3}(\frac{7\pi}{2},\frac{\pi}{2}%
)$\textbf{\ \ \ } & $3t_{m}$ & $\ \ 7.5\mu s$ & $\ $\\\cline{1-4}%
$e^{+i\pi/4\sigma_{z,2}}$ & $U_{I}^{2}(\frac{\pi}{2},\frac{\pi}{2})U_{I}%
^{2}(\frac{\pi}{2},0)U_{I}^{2}(\frac{7\pi}{2},\frac{\pi}{2})$ & $3t_{m}$ &
$\ \ 7.5\mu s$ & \\\cline{1-4}%
$e^{-i\pi/4\sigma_{y,3}}$ & $\ U_{I}^{3}(\frac{7\pi}{2},\frac{\pi}{2}%
)$\textbf{\ } & $t_{m}$ & $\ \ 2.5\mu s$ & \\\hline
\end{tabular}

\bigskip

\textbf{Table III.} Cases of three $^{171}$Yb$^{+}$ in a linear trap with the
largest $J$ (in our calculation) with respect to a certain inter-ion distance
$h$, where $W$ is the trap frequency, $\partial B/\partial z$ denotes magnetic
field gradient.%

\begin{tabular}
[c]{|c|c|c|c|c|c|}\hline
$h(\mu m)$ & $\ W(MHz)$ & $\partial B/\partial z(T/m)$ & $\ \varepsilon_{\max
}$ & $\ J(kHz)$ & $\ J_{13}(kHz)$\\\hline
$2$ & $1.77\times2\pi$ & $\ \ 750$ & $0.0276$ & $1.12\times2\pi$ &
$\ 0.794\times2\pi$\\\hline
$3$ & $0.966\times2\pi$ & $\ \ 300$ & $0.0271$ & $0.605\times2\pi$ &
$\ 0.429\times2\pi$\\\hline
$4$ & $0.628\times2\pi$ & $\ \ 150$ & $0.0263$ & $0.359\times2\pi$ &
$0.254\times2\pi$\\\hline
$5$ & $0.449\times2\pi$ & $\ \ 100$ & $0.0289$ & $0.311\times2\pi$ &
$0.220\times2\pi$\\\hline
$6$ & $0.342\times2\pi$ & $\ \ 50$ & $0.0218$ & $0.134\times2\pi$ &
$0.0952\times2\pi$\\\hline
\end{tabular}

\newpage

\newpage
\end{document}